\documentclass{symmetry12}

\begin{document}\setcounter{page}{111}

\FirstPageHeading{Kiselev}

\ShortArticleName{Towards an axiomatic noncommutative geometry of quantum space\/-\/time}

\ArticleName{Towards an axiomatic noncommutative\\ geometry of quantum space and time}

\Author{Arthemy V. KISELEV}
\AuthorNameForHeading{A.V. Kiselev}
\AuthorNameForContents{KISELEV A.V.}
\ArticleNameForContents{Towards an axiomatic noncommutative geometry of quantum space and time}

\Address{Johann Ber\-nou\-lli Institute for Mathematics and Computer Science, \\
University of Groningen, P.O.~Box 407, 9700~AK Groningen, The Netherlands}
\Address{Institut des Hautes \'Etudes Scientifiques, Le~Bois\/-\/Marie~35, Route de Chartres, 91440~Bures\/-\/sur\/-\/Yvette, France}
\EmailD{A.V.Kiselev@rug.nl}

\Abstract{By exploring a possible physical realisation of the geometric concept of noncommutative tangent bundle, we outline an axiomatic quantum picture of space as topological manifold and time as a count of its reconfiguration events.}



\noindent%
This is a physics\/-\/oriented companion to the brief communication~\cite{Kiselev:SQS11} and its formalisation~\cite{Kiselev:JGP12}; 
we analyse a possible physical meaning of the notions, structures, and logic in a class of noncommutative geometries considered therein, see also~\cite{Kiselev:KontsevichCyclic,Kiselev:OlverSokolovCMP1998}.
We now try to formalise the geometry of such intuitive ideas as space and time, aiming to recognise further such phenomena of Nature as the mass and gravity (here, dark matter and vacuum energy) and Hubble's law.
Affine Lie algebras, the definition of real line~$\mathbb{R}$ via~$0$, $1$, addition, and bisection, and 
Vo\-ro\-no\"\i\ diagrams play key r\^o\-les here. 

This text itself is a part of the essay~\cite{Kiselev:Noel12}: let us first describe the configuration of physical vacuum, while in the second half of~\cite{Kiselev:Noel12} we study the admissible means and rules of coding sub\/-\/atomic particles and explore the ways of the particles' (trans)\/formations, reactions, and decays. In particular, we then address the mass endowment mechanism, generation of (anti)\/matter and annihilation, CP\/-\/symmetry violation, three lepton\/-\/neutrino matchings, spin, helicity and chirality,
electric charge and electromagnetism, as well as the algorithms underlying the 
weak and strong processes. The goal which we set for a semantic analysis of the postulates and their implications within the algebra and calculus in~\cite{Kiselev:SQS11,Kiselev:JGP12} is the construction of an elementary, toy model unifying the four fundamental interactions. 
We agree that the potential of our topological and combinatorial picture to explain or propose possible (dis)\/verifying experiments is not still the required ability for a model to predict.

In fact, we only discuss a possible physical sense of axioms and operations or deductions which are admissible in the chosen setup. Still, our synthesis may be not a unique way to relate this mathematical formalism to Nature.


\begin{remark}
We attempt to identify and describe the physics which not necessarily~\textsl{is}. We now sketch the processes and motivate their laws which could be dominant in the early Universe only. 
Alternatively, it may happen that these processes are realised nowadays (or are presently registered as signals which were emitted from afar in our remote past) only under very restrictive hypotheses about the local space\/-\/time geometry, e.g., near a black hole or near its singularity. 
Nevertheless, we develop the formalism in a hope that it does render 
the quantum structure of the Universe at Planck scale.
\end{remark}

Our main message is this: It may be that at the Planck scale, the geometry of this world is disappointingly simple\footnote{ This is likely if we recall how Science gradually cast away various essences such as the phlogiston (though as an abstract principle it was useful for the technology of steam engine), \ae{}ther (to which we owe the radio and knot theory, recalling that W.~Thomson's vortex rings preceded Bohr's planetary model of atom), or the long\/-\/range gravity force (which remains helpful for navigation in the Solar system).}
because
\begin{itemize}
\item it does not refer 
to the \textsf{diffeo}\/-\/structure of the visible space, i.e., to its locally vector space organisation with velocity along piecewise\/-\/smooth trajectories and their length, and with smooth transition functions between charts in the atlas for that manifold; instead, the events occur in the Universe, which does not amount to the visible space, by using its much more rough \textsf{homeo}\/-\/structure of topological manifold with continuous transition functions, whereas the incidence relations between points along continuous paths replace the obsolete notions of length and speed;
\item the Universe consists of \textsl{naught} but the \textsf{homeo}\/-\/class space itself and the information which it carries or is able to carry; the fact of existence, behaviour, and known forms of the interaction between 
particles refer to the locally available information (in particular, stored in a single point by using a local modification of the topology); the presence of gauge degrees of freedom at each point of the \textsf{diffeo}\/-\/class space is the manifestation of its own \textsf{homeo}\/-\/structure; the gauge transformations are performed pointwise, either entirely independently at different points or in no more than a (piecewise) continuous way, whence an attempt to bind Nature with their differentiability --in order to introduce the gauge connections by taking derivatives of arbitrary functions-- is an \textsl{ad hoc} assumption of the objects' description.
\end{itemize}
Indeed,
%
let us notice that the idea of a connection in a principal fibre bundle appeals to the (existence of) structures, and to their values outside that point, which therefore requires the existence of other points. Moreover, the gauge freedom of any kind allows, in its basic formulation~\cite{Kiselev:Okun}, pointwise\/-\/uncorrelated gauge transformations of the field of matter. Consequently, the postulate that \textsl{length} is defined and hence the distance between space\/-\/time points and between fields can be measured, giving rise to the construction of derivatives, and the postulate that the pointwise\/-\/uncorrelated values glue to a (piecewise\/-)\/smooth local section are the act of will, i.e., an \textsl{ad hoc} assumption in a description to which Nature is indifferent (e.g., see~\cite[Eq.~(4,1)]{Kiselev:LLIV}).
We remark separately that the operations which are recognised as gauge transformations but which stem from the presence and structure of space itself (e.g., local homeomorphisms of topological spaces) are at least but also no more than continuous, see sec.~\ref{RemSU2Illusion} on p.~\pageref{RemSU2Illusion}. We conclude that 
local processes can not be governed by smooth gauge theory (or only by it; in particular, gauge connection fields did not exist at the moment of the Big Bang).

\smallskip
If this is indeed so, the rules of Nature's behaviour are the arithmetic --comparison or addition of topologically\/-\/defined integer numbers-- and the associative algebra of gluing or splitting words written consecutively in its alphabet(s), and the coding of such topological objects as walks and cycles or knots.

\section{The two avatars}
This Universe is a topological space. In the beginning, its topology was trivial: $\mathcal{T}_0=\{\varnothing,\text{Universe}\}$.\ Nowadays it is not Hausdorff 
so that there are points which we can not tell one from another.\ Modifications of the topology~$\mathcal{T}$ are also possible.\footnote{We claim that exclusions of sets from the list~$\mathcal{T}$ and re\/-\/inclusions of the information about such sets provide the mass endowment mechanism and formation of the black holes' singularities.}

Within the regions of \textsl{vacuum} where the topology allows one to distinguish between points, the Universe is endowed simultaneously with two structures: one is the \textsf{homeo}\/-\/class structure of topological manifold with \textsl{contimuous} transition functions between coordinates (those form continuous nets on the charts); the other is the \textsf{diffeo}\/-\/structure of smooth manifold such that the transition functions are \textsl{smooth} and local coordinates form the smooth nets.

The Universe co\/-\/exists in its \textsf{homeo} and \textsf{diffeo} avatars. The \textsf{homeo} structure is the \textsl{quantum world}; it carries the information about the geometry and about the types, formation, and actual existence and states of the particles. The laws of fundamental interactions between particles retrieve and process that information, thus determining the processes that run at the quantum level. Each particle or any other object in the quantum, \textsf{homeo}\/-\/class geometry has a continuous world\/-\/line.

Our conscience percepts the Universe and events in it at the macroscopic level using the smooth, \textsf{diffeo}\/-\/class geometry, that is, by understanding of the local charts as domains in a vector space over~$\mathbb{R}$ with the usual arithmetic of vectors. The notion of \textsl{length} is defined in the macroscopic world.\footnote{By convention, a length scale is \textsl{macroscopic} if the typical distances considerably exceed the electric\/-\/charge diameter of proton, which is approximately 1~\textsf{fm}~$=10^{-15}$~\textsf{m}.} This notion allows us to measure local macroscopic distances by using rigid rods and also measure local time intervals between events by employing the postulate of invariant light speed, that is, by using derivatives of the former equipment. With the help of rigid rods and light, we introduce the macroscopic notions of instant velocity and define the nominal concept of a smooth trajectory, not referring it to any material object but only to the local properties of the smooth macroscopic space\/-\/time. The transition between macroscopic charts with smooth coordinates in the \textsf{diffeo}\/-\/class space\/-\/time are governed by the Lorentz transformations. The topology of macroscopic space\/-\/time outside particles and black hole singularities is induced by the (indefinite) metric in inertial reference frames.

The tautological mapping from the quantum, \textsf{homeo}\/-\/class world to the macroscopic, \textsf{diffeo}\/-\/class realisation of the Universe is continuous but not a homeomorphism; by construction, it can not be an isometry. Under this tautological mapping, the information which is realised by the quantum geometry takes the shape of particles in the macroscopic world; however, a part of this information is lost along the way due to the introduction of length (more precisely, of Lorentz' interval): there are topologically\/-\/nontrivial quantum objects --in fact, a whole dimension-- which acquire zero visible size in all reference frames.

On top of that, because the composition of (1) local homeomorphisms from the standard domains to the charts of the \textsf{homeo}\/-\/class topological manifold with (2) the tautological mapping is only continuous at all points of the Universe, the images of points and point particles in its visible, \textsf{diffeo}\/-\/class realisation are observed by us as if they are in a perpetual inexplicable ``motion.'' Namely, as it often happens with legal documents, no other rights may be derived from the statement that the composition is continuous: the pledge is to take points from nested sets in the atlas~$\mathcal{T}$ to near\/-\/by points as we see them, but the continuous mapping does not presume that we, upon our own initiative, shall apply our notion of length to some continuous curves connecting those images. In effect, inertial trajectories of material objects in the quantum world are continuous but nowhere differentiable.\footnote{This creates the classical antagonism between Red and White: namely, Norm is firm, straight, and always all right while Shake is indecisive and trembling. Only we cannot tell who is who in this Universe.}
The visible world\/-\/lines are at most $(c,1)$-\/Lipschitz, where $c$~is the speed of light and the power~$1$ states that no material object is allowed to run out of the light cone of its future.

\begin{example}
Suppose that we know (setting aside all the subtleties related to the act of measurement) that a quantum object --e.g., a marked point of it where all the mass or all the charge is contained-- is located now at a given point and moreover, it does not move with respect to other points according to the incidence relations between points in a Hausdorff topology. Nevertheless, we may not know its visible instant velocity because that notion refers to the limit procedure in a vector space and hence is not applicable. 
\end{example}

\begin{example}
Likewise, choose an inertial reference frame and consider a situation when a domain in quantum space is homeomorphic to a domain within a crystal structure~\cite{Kiselev:Humphreys,Kiselev:KacInfDim}. Suppose that a vertex of the lattice decides to visit its neighbour and thus goes along the edge connecting them. Not only its visible initial location was non\/-\/constant in time and the initial instant velocity undefined, but this will remain so at all points of the continuous, nowhere\/-\/differentiable image of the trajectory along the edge; the journey will end at an unpredictable location of the endpoint with undefined terminal velocity. We conclude, referring again to the Red\/--\/White antagonism, 
that it is impossible to tell which of the two worlds, topological \textsf{homeo} or smooth \textsf{diffeo}, is straight and which is shaking.
\end{example}

For the same reason, provided that we postulate a point particle's \textsl{visible} instant velocity (irrespective of its actually on\/-\/going displacement with respect to the topologically\/-\/neighboured points in the quantum world), we may not determine at which point of the visible macroscopic space it is located. (Let us remark that the above examples and reasonings are not applicable to the propagation of light which can not stay at rest with respect to the incidence relations.) The balance of resolution for the location of a material quantum object in the smooth space at a given time and for its momentum is determined by the Heisenberg uncertainty principle.\footnote{Note that we may not track the behaviour of ``empty'' points of the quantum space if they are not referred to by any material object located there; consequently, we do not attempt to introduce a ``temperature'' of the vacuum.}
Simultaneously, the propagation of a point particle from a given point to a given endpoint along an \textsl{a priori} unknown continuous trajectory in the visible, \textsf{diffeo}\/-\/class world is the cornerstone of the concept of Feynman's path integral.

We now propose to abandon the futile attempts to measure or approximate the undefined notions but study the interactions between quantum objects by referring their laws to the \textsf{homeo}\/-\/class geometry of the Universe. Let us remember that the difficulties and uncertainties which we gain --when measuring length and calculating derivatives such as the velocity-- in a description of the quantum processes do not stem from their true nature. Integrating empirically the laws of its evolution, the Universe stays, and will stay forever indifferent to the fact that we can not grasp all its details at once, since we ourself first proclaimed our intention to take proportions with respect to the standard metre instead of inspecting the topological invariants of phenomena.

\begin{corollary}
The processes in the quantum, \textsf{homeo}\/-\/class (locally) Hausdorff topological manifold without length can not be adequately described by (the geometry of partial) differential equations (c.f.~\cite{Kiselev:GDE12,Kiselev:Olver}). On the other hand, the construction of $\sigma$\/-\/algebras associates the \textsl{measure} in its true sense with sets of points but not with distances between points; consequently, \textsl{integral} equations could be more relevant. 
\end{corollary}


\section{The time phenomenon}\label{SecTime}
There are at least two ways to understand what the \textsl{time} is in context of a paradoxal observation by our conscience that everything in this world is staying perpetually in the \textsl{present}.

A realist approach to the notion of time postulates the existence of a full\/-\/right uncompactified dimension with a reasonable topology of the resulting space\/-\/time. One then operates with the count of time by using the invariant Lorentz interval, light cones of the past and future, and world\/-\/lines of material objects. An inconvenience of this approach is that, in order to maintain the everlasting presence, the visible world must unceasingly glide along the time direction, i.e., 
to keep in the same place, it takes all the running one can do.
Note that under Lorentz' transformations the local observer's time can be bent towards another observer's space and \textsl{vice versa} but in earnest the time can not be swapped with any spatial direction.

The concept of a $(3+1)$\/-\/dimensional smooth or topological manifold into which the time is incorporated \textsl{a priori} contains the following logical difficulty. An infinitely\/-\/stretching absolutely empty, flat Minkowski space\/-\/time $\bigl(\mathbb{R}\times\mathbb{R}^3,({+}{-}{-}{-})\bigr)$ without a single object in it would exist \textsl{forever}. In our opinion, there is no time at all in that empty world: the cups, tea, and bread\/-\/and\/-\/butter always remain the same, so it is always six o'clock. Nobody counts to the Time hence 
time does not count.

\begin{definition}
We accept that the time is a count of reconfiguration events in this Universe; such events are, for example,\footnote{Were the Universe truly smooth, Poisson, and possess the Hamiltonian functional, then the time by definition would be taking the Poisson bracket with such master\/-\/functional of the current state; a weakened and much more likely formulation is the generation of time by events of evaluating binary operations at locally defined Hamiltonian functionals that correspond to separate particles, c.f.~\cite{Kiselev:JGP12}.}
the reconfigurations of geometry (i.e., an act of modification in the topology) or the operation of an algorithm that transfers information over points, creating an event of output statement by processing the local configuration of the Universe in its input (such is the propagation of light).
\end{definition}

Thus, events create time. Events which do not reconfigure the Universe (e.g., a correct statement that for a topologically\/-\/admissible arc connecting point~$\underline{0}$ to point~$\underline{1}$ there is a null path running from~$\underline{0}$ to~$\underline{1}$ and then back again along the same arc) \textsl{do not} express the count of time (although a \textsl{verification} of such statement by using light signals 
does take and hence creates time).

The notions of recorded past and expected future are derived from the relation of order in the count of events by an appointed observer; let us remember that an opinion of another observer about the order of events could be different.

\begin{example}
Consider the reconfiguration of the Universe produced by a trip of Chapeau Rouge from point~$\underline{0}$ to point~$\underline{1}$ along a continuous arc connecting them in a coordinate chart of the \textsf{homeo}\/-\/realisation of the Universe. This amounts to the input information that the two endpoints, the arc, and Chapeau Rouge exist, that the available choice of topology confirms that the path is continuous, and to the work of the algorithm the negates the already passed points and thus prescribes the admissible direction to go further.

In absence of length and in absence of any devices at the observer's disposal, the time is discrete: it is counted by the events (1) Chapeau Rouge is at the starting point; (2) Chapeau Rouge has reached the endpoint. 

The observer can grind the time scale by recursively installing the intermediate checkpoints somewhere in between the points which are already marked; this is done by using the incidence relation for points on the continuous path and does not refer to the notion of length (in fact, it refers to the definition of real numbers by using $0$, $1$, addition, and bisection). The limiting procedure makes the count of time continuous.

It is the postulate of invariant light speed which endows the Universe with its local smooth structure (``twice earlier $\Longleftrightarrow$ twice closer''). The light automaton is programmed to choose the next point by processing 
the information about earlier visited points and creates an event of specific type; the principle is that all observers accept its performance identically. By using the bisection method, we first mark the midpoint~$\underline{{}^{1}/{}_{2}}$ on the chosen curve and replicate the automaton $\underline{0}\to\underline{1}$ to the automata $\underline{0}\to\underline{{}^{1}/{}_{2}}$ and $\underline{{}^{1}/{}_{2}}\to\underline{1}$; then we declare that the old automaton counts the unit step of time and each of the new automata counts one half. The recursive process and the limiting procedure create the smooth structure of space\/-\/time for a given observer.\footnote{To use light as the pacemaker of the clock, we ought to describe first what a photon is; to do that we operate in~\cite{Kiselev:Noel12} with the \textsf{homeo}\/-\/class geometry. We also notice that \textsl{before} the emission of the first photon in history of the Universe (or before creation of any other massless particles which travel with the same invariant speed~$c$), the time had been counted by using events of other origin; we argue that such events were the reconfigurations in the topology~$\mathcal{T}$.}

The inconvenience is that this smooth structure is not applicable to material objects which are known to travel slower than light; in order to monitor the steady progress of Chapeau Rouge on her way from~$\underline{0}$ to $\underline{1}$, one must use as many light signals as there are checkpoints installed along the path. Even if the \textsl{energy} emitted by the new, ``shorter\/-\/range'' automata drops at the moment of each replication, the total energy which one has to spend in the continuous limit is either null or infinite; the first option is useless because it does not communicate any information to the observer; the second option is not impossible if the Universe is infinite and the observer agrees to waste a finite fraction of this world, still it is impractical.
\end{example}

In the next section we introduce a possible local topological structure of the \textsf{homeo}\/-\/class realisations of the Universe. If one feels it necessary to multiply dimensions, then we advise to let a macroscopic observer view the construction from a chosen inertial frame; the time direction is then locally decoupled to the real line~$\mathbb{R}$. At the end of the next section, the structure of the macroscopic images of domains under the tautological mapping from \textsf{homeo} to \textsf{diffeo} is then recalculated to all other inertial reference frames by using Lorentz' transformations. Let us only remark that the ``smooth time'' parameter is introduced in the \textsf{diffeo}\/-\/class world in order to legalise the limiting procedures such as the correlation of arbitrarily small length and the speed of light in the local vector\/-\/space organisation of space.

\section{Local configuration of quantum space}
Now we introduce the local topological configuration of empty quantum space, that is, \textsl{vacuum} away from the singularities of black holes. In technical terms, we define the admissible local structure by taking ``as is'' or via self\/-\/similar continuous limit the lattice of affine Lie algebra (primarily using the root systems~$A_3$, $B_3$, or~$C_3$ of simple complex Lie algebras) and thus consider a truncated Kaluza\/--\/Klein model without the Minkowski dimension of Newtonian time but instead, with a topology brought in by hand (though it is equivalent to a standard one for each continuous limit); we then analyse the origins of Hubble's law.

To describe a domain in quantum, or \textsl{quantised}, space we first consider Euclidean space~$\mathbb{E}^3$ containing the affine lattice generated by the irreducible root system~$A_3$, $B_3$, or~$C_3$ (see Remark~\ref{RemContinuousLim} on p.~\pageref{RemContinuousLim}). Let us denote by ${\vec{\mathrm{x}}}_i\equiv{\vec{\mathrm{x}}}_i^{\,+1}$ the generators of the lattice at hand and by ${\vec{\mathrm{x}}}_i^{\,-1}$
their inverses (so that the paths $\underrightarrow{{\vec{\mathrm{x}}}_i\cdot{\vec{\mathrm{x}}}_i^{\,-1}}=
\underrightarrow{{\vec{\mathrm{x}}}_i^{\,-1}\cdot{\vec{\mathrm{x}}}_i}=1$ end at the point where they start). Each lattice determines the tiling of space~$\mathbb{E}^3$, its vertices and edges constituting the $1$-\/skeleton of the CW-\/complex with trivial topology (see Remark~\ref{Kiselev:RemLoll} on p.~\pageref{Kiselev:RemLoll}). We let a finite domain in~$\mathbb{E}^3$ with a given configuration of vertices and the adjacency table of the lattice be the \textsl{spatial component} of a prototype domain in the discrete quantum space.

Second, we take the product~$\mathbb{E}^3\times\mathbb{E}^2$ of space with a two\/-\/plane into which we place the circle~$\mathbb{S}^1$ passing around the origin. Viewing the circle as an oriented one\/-\/dimensional topological manifold, we create an extra, compactified dimension in the local quantum geometry. Namely, to each vertex of the prototype domain we attach the \textsl{tadpole}~$\mathbb{S}^1$, i.e., the edge that starts and ends at the same point and loops in the extra dimension (outside the old~$\mathbb{E}^3$). By convention, we denote by $\mathbb{S}^1\equiv\mathbb{S}^{+1}$ the tadpole walked counterclockwise with respect to the standard orientation of~$\mathbb{E}^2$ and by $\mathbb{S}^{-1}$ the reverse, clockwise cycle.

\begin{remark}
Under the tautological mapping of the quantum world to the \textsf{diffeo}\/-\/class visible world, the tadpoles are assigned zero length because the distance between their start-{} and endpoints vanishes for each of them.\footnote{We note that the notion of length \textsl{is} applicable to the generators $\mathbb{S}^{\pm1}$ of such null vectors but it is not applicable to the edges~${\vec{\mathrm{x}}}_i^{\,\pm1}$ in space: their length is undefined because the homeomorphisms from domains in~$\mathbb{E}^3$ containing the lattice to the spatial counterparts of the prototype domains are not fixed but can change with time.}
We conclude that the entire compactified dimension is invisible to us; this is why the tautological mapping between the \textsf{homeo}-{} and \textsf{diffeo}\/-\/realisations of the Universe is not bijective: it compresses one extra dimension at each point to a null vector.\footnote{We also notice that the generators~$\mathbb{S}^{\pm1}$ encode nontrivial walks in quantum space but produce no visible path which would leave a single point in the macroscopic world; we postulate that the contours~$\mathbb{S}^{\pm1}$ determine the electric charge~$\pm e$, 
see~\cite{Kiselev:Noel12}.}

Namely, passing to the additive notation $\bigl(({\vec{\mathrm{x}}}_i,\vec{\mathrm{d}}),\pm\bigr)$ instead of the multiplicative alphabet $\bigl(({\vec{\mathrm{x}}}_i^{\,\pm1},\mathbb{S}^{\,\pm1}),\cdot\bigr)$, that is, viewing the letters as vectors in Euclidean space but not as the shift operators and introducing the \text{null vector}~$\vec{\mathrm{d}}$, we recover the standard description of the affine basis for the Kac\/--\/Moody algebra at hand; clearly, the length of the null vector equals zero.
Recall further that the circle~$\mathbb{S}^1$ is the total space in a double cover over the real projective line $\mathbb{RP}^1\simeq\mathbb{S}^1/{\sim}$; one full rotation~$\mathbb{S}^{+1}$ corresponds\footnote{We introduce a separate notation~${\vec{\mathrm{t}}}$ for one rotation along the projective line anticipating its future application in the description of ``building blocks'' for strong interaction 
and also its possible use in the study of the quantum Hall effect.} to running along the projective line twice: $\mathbb{S}^1=\overrightarrow{\mathrm{tt}}$ and $\mathbb{S}^{-1}=\bigl(\overrightarrow{\mathrm{tt}}\bigr)^{-1}$;
the double cover over~$\mathbb{RP}^1$ is then responsible for the familiar coefficient `$2$' in front of the null vector~$\vec{\mathrm{d}}$. 
\end{remark}

\begin{remark}
The vertices of the CW-\/complex are the \textsl{quanta} of space; there is a deep logical motivation for their existence. Namely, by assembling to one vertex a continuum of physical points within a domain which is dual to the set of neighbouring vertices in the lattice, Nature replaces the \textsl{continuous} adjacency table between points to a \textsl{finite}, lattice\/-\/dependent table so that there are only finitely many neighbours of each vertex and hence a finite local configuration of information channels.

In conclusion, space is continuous but the Universe operates with quantum phenomena in it, thus achieving a great economy in the information processing.
\end{remark}

\begin{remark}
The tadpole~$\mathbb{S}^1$ at a vertex of quantum space is an indexed union $\bigcup_{i\in\mathcal{I}}\mathbb{S}_i^1$ of tadpoles referred by~$i$ to an indexing set~$\mathcal{I}$ of points in the quantum domain which is marked by the vertex. Typically, this set is at least countable, $\mathcal{I}\supseteq\mathbb{Z}$; one could view it as an enumerated set of binary approximations for points in that domain (here we use the auxiliary metric in~$\mathbb{E}^3$); we emphasize that by choosing the indexing set in this way we endow it with \textsl{order}, c.f.~\cite{Kiselev:Noel12}. 

This convention allows us to handle infinitely many tadpoles attached to an everywhere dense set of the quantum domain by ascribing a different statistics to a unique tadpole which is attached to the vertex which marks that domain.
\end{remark}

\begin{note}
We postulate that the spatial edges ${\vec{\mathrm{x}}}_i^{\,\pm1}$ of the lattice are \textsl{fermionic} so that no such edge can be walked twice in the same direction by one path; a path can run twice along the same edge only in the opposite directions. Note that different paths can go independently in the same direction along a common edge; we also notice that a path can run many times through a vertex, approaching it each time by a different edge in its adjacency table.

Unlike it is with spatial edges, the tadpoles $\mathbb{S}^{\pm1}$ attached to the vertices are \textsl{bosonic} so that paths can rotate on these caroussels any finite number of times in any direction (which does not really matter because the overall difference $\sharp\mathbb{S}^1-\sharp\mathbb{S}^{-1}$ of positive and negative rotation numbers is constrained by the value of electric charge of the particle encoded by the path). However, let us remember that in earnest we are dealing with ordered infinite sets $\{\mathbb{S}^{\pm1}_m$, $m\in\mathbb{Z}\subseteq\mathcal{I}\}$ of \textsl{fermionic} tadpoles brought to the marker of a quantum domain; in the continuous limit of a quantum tiling, this set spreads over the domain~--- one fermionic tadpole per each indexed point.
\end{note}

So, let us recall that the part of a lattice in a domain of~$\mathbb{E}^3$, with tadpole attached to each vertex, is \textsl{discrete}. We say that the original fermionic lattice with bosonic tadpoles is the \textsl{quantum space}; in what follows we formalise the geometry of elementary particles in terms of the \textsl{alphabet} $\mathfrak{A}=\bigl(({\vec{\mathrm{x}}}_i^{\,\pm1},\mathbb{S}^{\pm1}),\cdot\bigr)$. 

The standard bisection technique (see sec.~\ref{SecTime}) allows us to convert the discrete tiling to its continuous limit in which the topology is inherited from the adjacency table of the affine lattice (the neighbourhoods 
in the Vo\-ro\-no\"\i\ diagram
are the duals of adjacent vertices' configuration in the spatial, $\mathbb{E}^3$-\/tiling component of the CW-\/complex); the limit topology is locally equivalent to the product topology for~$\mathbb{S}^1$ and Euclidean space~$\mathbb{E}^3$; the orientation field for~$\mathbb{S}^1$ over~$\mathbb{E}^3$ is continuous.

\begin{definition}
The self\/-\/similar limit of the discrete structure in a domain of quantum space is a \textsl{domain} in the \textsf{homeo}\/-\/class realisation of the Universe.
\end{definition}

\begin{remark}
The introduction of a continuous field of fermionic lattice generators ${\vec{\mathrm{x}}}_i^{\,\pm1}$ and fermionic loops $\mathbb{S}^{\pm1}$ or ${\vec{\mathrm{t}}}^{\,\pm1}$ over \textsl{each} point of continuous space, which we have performed here in full detail, is the \textsf{homeo}-\/class analog of the \textsl{noncommutative tangent bundle} over the smooth visible realisation of the Universe, see~\cite{Kiselev:JGP12}.
\end{remark}

Quantum space is discontinuous; in sec.~\ref{SecTime} we argued that a verification of the continuity for its self\/-\/similar limit requires the expense of infinite energy whenever one attempts to monitor a steady motion of a material object travelling slower than the speed of light and for that purpose encodes the object's path by the alphabet $\mathfrak{A}_\infty=\bigl((\tfrac{1}{2^n}{\vec{\mathrm{x}}}_i^{\,\pm1},\mathbb{S}^{\pm1};\ n\in\mathbb{N}\cup\{0\}),\cdot\bigr)$. However, a motivation why the limit should nevertheless be studied --and is more than a mathematical formality-- is as follows. Namely, a \textsl{continuous} coding of points in space by using binary arithmetic permits us to consider continuous paths --in particular, closed contours,-- not referring them to a specific lattice. Indeed, our ability to describe and handle such contours does not imply that any material object is actually transported along those paths; hence energy is not spent but the drawn figures, and homotopies of these images in space, do encode information: a generic continuous path is an infinitely\/-\/long cyclic word written by using infinitely\/-\/short letters of the alphabet~$\mathfrak{A}_\infty$. The massless chargeless contours propagate freely in \textsf{homeo}\/-\/class domains until a very rare event of their disruption and weak interaction with other material objects. 
However, this is only a part of the story.

\subsection{The $U(1)\times SU(2)$-\/picture}\label{RemSU2Illusion}
First, let us notice that there is no marked origin in the affine lattice and therefore it acts on itself by finite shifts. Note further that this action is topological: it appeals to the incidence relations between vertices but not to the smooth, local vector\/-\/space organisation of~$\mathbb{E}^3$.

Having placed the affine lattice in~$\mathbb{E}^3$, one could --by an act of will to which Nature is indifferent-- extend the algebra of finite shifts to the space of homeomorphisms of~$\mathbb{E}^3$, i.e., the local action of space upon itself by a continuous field of translations. Moreover, by compactifying space to $\mathbb{E}^3\cup\{\text{pt}\}\simeq\mathbb{S}^3$, one extends this action to homeomorphisms of the three\/-\/sphere. By yet another misleading isomorphism $\mathbb{S}^3\simeq SU(2)$ --which is given, e.g., by the Pauli matrices-- one is tempted to conclude that
\begin{enumerate}
\item the complex field~$\mathbb{C}$ is immanent to static geometry of the Universe, and
\item the freedom of appointing for reference point any vertex in the affine lattice, now realised as a set of points inside~$SU(2)$, means the introduction of the $SU(2)$-\/principal fibre bundle over the space\/-\/time.
\end{enumerate}
Yet even more: though the pseudogroup of local homeomorphisms of space states that the field of pointwise\/-\/defined shifts is continuous, it is postulated
that this deformation field is smooth, hence there exist derivatives of local sections for the principal fibre bundle.
This pile of \textsl{ad hoc} conventions delimits the smooth complex $SU(2)$-\/gauge theory of weak interaction~\cite{Kiselev:Okun}.

Likewise, each tadpole's circle~$\mathbb{S}^1$ carries the gauge freedom of marking a starting (hence, end-{}) point on it and also it can be subjected to an arbitrary homeomorphism (not necessarily a diffeomorphism), which leaves the tadpoles~$\mathbb{S}^{\pm1}$ intact. The choice of marked points is made \textsl{pointwise} at vertices of the lattice (or at all points of continuous space if we deal with the limit) --- without any idea of smoothness superimposed to continuity. Now we note another misleading isomorphism~$\mathbb{S}^1\simeq U(1)$, which also tempts one to introduce complex numbers in the static quantum geometry.

Summarising, we see that electroweak phenomena could be quantum space in disguise.

\subsection{Hubble's law}
Second, let us recall that there are exactly three canonical tilings of Euclidean space.

\begin{remark}\label{RemContinuousLim}
The three irreducible tilings of space are determined by simple root systems~$A_3$, $B_3$, and~$C_3$. They have equal legal rights in the geometric construction, still we believe that the tetrahedral tiling which corresponds to~$A_3$ dominates over the two others whenever one is concerned with the symmetry and stability of particles whose contours are encoded by words written in these root systems' alphabets (see~
\cite{Kiselev:Noel12}). Thus, more symmetric particles are more stable.

We recall that the adjacency tables for vertices are different for the three irreducible lattices in~$\mathbb{E}^3$ so that the local 
configurations of information channels between points in the continuous limits are also different; the three continuous versions of one space differ by the algorithms of processing locally available information.\footnote{Recall that the two gentlemen of Verona could embark and sail to Milan with the morning tide; alternatively, they could take a train, fly in an airplane, or go by car. Their route organisation would have been different in these four cases, yet the starting-{} and end\/-\/points coincide; the four paths are integrated by  rotation of screw or wheel.}
However, the limit topologies are equivalent in a sense that a continuous path in one picture stays continuous in any of the other two; the transliteration of a continuous path then amounts to a \textsl{second order phase transition} when the object stays identically the same but the underlying crystal structure changes.\footnote{We expect that the transliterations --from one alphabet to another-- of cyclic words encoding the contours 
whose meaning is a 
chargeless spin\/-$\tfrac{1}{2}\hbar$ particle explain the known neutrino oscillation.}

In the sequel, we prefer to operate locally on the affine lattice~$A_3$ yet we allow a formal union of the three irreducible alphabets in the fibre of the noncommutative tangent bundle over each point of space. We view the irreducibility as the mechanism which holds space from slicing to lower\/-\/dimensional components; because of that, we shall not consider the reducible cases $A_1\oplus A_1\oplus A_1$, $A_1\oplus A_2$, $A_1\oplus B_2$, and~$A_1\oplus G_2$. We also emphasize that we always preview a possibility of taking the continuous limit in mathematical reasonings but we let the space be quantised by the edges of the graph, i.e., by the $1$-\/skeleton of the CW-\/complex.
\end{remark}

Viewing the world as it is (e.g., compared with the multiplicative structures in~\cite{Kiselev:JGP12}), we have to admit that a perfectly ordered life inside a Kac\/--\/Moody algebra is an inachievable ideal.
In practice, the $1$-\/skeleton of the CW-\/complex experiences an everlasting reconstruction; this is why up to this moment we have not described the attachment algorithm or transition mappings between overlapping quantum domains; they just attach as graphs and the verity is that the CW\/-\/complex is globally defined --- it \textsl{is} space in which the Universe exists.

A possible mechanism of the perpetual modification in the graph's local topology (but not in the triviality of topology of the CW\/-\/complex) is that Natura abhorret a vacuo. In its zeal to shake off its quantum discontinuity, Nature does attempt to perform the infinite bisection and construct the complete real line~$\mathbb{R}$ by using binary arithmetic. Let us recall that such recursive procedure replicates one unit\/-\/time light automaton to two automata plugged consecutively, one after another. But Nature unceasingly replicates each light automaton with its two copies that are \textsl{identical} to their sample. This leads to the observed proper elongation of space.\footnote{Notice that a release of energy in the course of edge decontractions (see sec.~\ref{SecMass}) is compensated with a simultaneous increase of the volume of continuous space so that the energy density remains constant; then, a part of this energy is being absorbed by the black holes or is radiated to the spatial infinity.
A simultaneous release of ever\/-\/growing amount of energy per unit\/-\/time  at the outer periphery of decontracting Universe (see next section) does not burn the objects inside it but it instead cools down to the present $2.725^{\circ}$~\textsf{K} of the cosmic microwave background radiation.}

Namely, within each fixed half\/-\/time\footnote{This time interval is counted by the local billiard clock --the edge itself-- that sends light signals to and fro the edge; this parameter can vary as the Universe grows older.} on\/-\/average one half of the actually available edges split in two new edges. (We remark that this division does not happen with the \textsl{contracted} edges, see next section; on the same grounds it is the edges but not vertices that split, for the latter could in fact be a superposition of many vertices according to the record of past modifications in local topology.) Each event of edge splitting creates a new vertex --the midpoint-- and fills in the adjacency table for it, connecting it by one edge with all vertices in the cells delimited by the splitting edge in their faces.

Also, a tadpole is attached to the new vertex within the compactified dimension.
We recall that the vertices label quantum domains in space so that the graph's adjacency table configures the domains' neighbours. We now note that the process of spontaneous edges' splittings roots in the conventional round\/-\/up $[n-\tfrac{1}{2},n+\tfrac{1}{2})\mapsto n$: the edge's midpoint is referred to \textsl{one} of the edge's endpoints~--- hence, the midpoint's fermionic tadpole is communicated to that endpoint. The splitting goes as follows: in terms of the order in~$\mathbb{Q}$ along the splitting edge $(n-1,n)$, its midpoint $\{n-\tfrac{1}{2}\}$ detaches from~$\{n\}$, proclaiming the existence of a separate quantum domain $[n-\tfrac{3}{4},n-\tfrac{1}{4})$ which becomes adjacent with $[n-\tfrac{3}{2},n-\tfrac{3}{4})$ and $[n-\tfrac{1}{4},n+\tfrac{1}{2})$; the round\/-\/up demarkation reproduces twofold at $\{n-\tfrac{3}{4}\}$ and $\{n-\tfrac{1}{4}\}$. The marker~$\{n-\tfrac{1}{2}\}$ of the new domain grabs --and endows with order the set of-- fermionic tadpoles 
attached to all the points $\{n-\tfrac{3}{4}+\tfrac{m}{2^\ell}\}$ which (by the values of $m,\ell\in\mathbb{N}\cup\{0\}$) locally get into the bounds $[n-\tfrac{3}{4},n-\tfrac{1}{4})$.

But because the light automata remain the same for the first and second fragments of the edge, each of them counts the propagation of light signal along each new edge as a unit\/-\/time event. Consequently, not only the Universe grows at its periphery, but a trip between distant objects all across the Universe takes more and more time.

\begin{corollary}[The Hubble law]
The Doppler\/-\/shift\/-\/measured velocity~$v$ at which distant material objects, locally staying at rest with respect to physical points, i.e., with respect to the incidence relations between points in quantum space, recede from each other is \textsl{directly proportional} to the proper distance~$D$ between these objects:
\[
v=H_0\cdot D,
\]
here $H_0$~is the Hubble constant \textup{(}now it approximately equals~$74.3\pm 2.1$\,\textsf{\textup{(km/s)\,/Mpc}}\textup{).}

Notice that the picture is uniform with respect to all observers associated with such objects anywhere in space.\footnote{A relative motion of the Milky Way with respect to the underlying quantum space structure is observed by the detection of Doppler's shift in the relic radiation at certain direction and its antipode.}
\end{corollary}

Thus, Hubble's law testifies steady self\/-\/generation of space due to which Cosmos obeys the principle ``twice farther, twice faster'' at sufficiently large scale. We conclude that we do hear the process of space expansion in the form of the cosmic microwave background radiation; we thus predict that the $1.873$\,\textsf{mm}\/-\/signal can not be altogether shielded by any macroscopic medium.

\section{The mass}\label{SecMass}
The crucial idea in our description of the geometry of vacuum --which is not inhabited yet by any particles-- is that contractions of edges in the graph are allowed (but highly not recommended unless possible consequences are fully understood). We emphasize that this does not require stretching, pulling, compressing, or any other forms of physical activity~--- an ordinary accountant with pencil, eraser, and access to the book~$\mathcal{T}$ with topology of the Universe can accomplish more epic deeds in the course of one reaction than Heracles did in his entire life.

\begin{definition}
The \textsl{contraction} of an edge is a declaration that its endpoints merge and there remains nothing in between (i.e., a tadpole is not formed); the respective ordered sets of fermionic 
tadpoles~$\mathbb{S}$ attached to the merging endpoints 
unite, preserving the tadpoles' directions and their ordering
(so that a path in positive or negative direction along either 
bosonic or fermionic understanding for the old tadpoles becomes the respectively directed path on the new one). The \textsl{decontraction} of a previously contracted edge is its restoration in between its endpoints which become no longer coinciding, and the
splitting of the indexed ordered tadpole sets between the two
endpoints.
\end{definition}

\begin{remark}
A contraction of edges in the $1$-\/skeleton of the CW-\/complex can force the formation of tadpoles from remaining edges: 
for 
example, 
two contractions which distinguish between Left and Right 
could lead to the CP\/-\/symmetry violation in weak processes (see~\cite{Kiselev:Noel12}).

Let us also notice that the on\/-\/going splitting of edges, which is responsible for the Hubble law, is a random 
decontraction process spread over quantum 
space.
\end{remark}

\begin{figure}[htb]
{\unitlength=1mm
\centerline{
\begin{picture}(105,30)(-5,-3)
\put(-3,0){\vector(1,0){13}}
\put(0,-3){\vector(0,1){13}}
\put(-2,-2){\vector(1,1){9}}
\put(11,-1){${\vec{\mathrm{x}}}$}
\put(7.5,7){$\vec{\mathrm{y}}$}
\put(-1,8){\llap{$\vec{\mathrm{z}}$}}
\put(25,0){\begin{picture}(40,20)
\put(0,0){\circle*{1}}
\qbezier[16](0,0)(16,0)(16,0)
\put(16,0){\circle*{1}}
\put(0,0){\line(0,1){16}}
\put(0,16){\circle*{1}}
\put(0,16){\line(1,0){16}}
\put(16,16){\circle*{1}}
\put(16,0){\vector(0,1){15.5}}
\put(-1,-3){\llap{$a$}}
\put(16.5,-4.2){$b$}
\put(-1,16.5){\llap{$d$}}
\put(16.5,16.5){$c$}
\put(15,11){\llap{$\vec{\mathrm{z}}$}}
\put(16,0){\vector(1,0){16}}
\put(32,0){\vector(1,1){8}}
\put(40,8){\vector(-1,0){16}}
\put(24,8){\vector(-1,-1){8}}
\put(24,-4){${\vec{\mathrm{x}}}$}
\put(37,2){$\vec{\mathrm{y}}$}
\put(32,9){${\vec{\mathrm{x}}}^{\,-1}$}
\put(21,2){$\vec{\mathrm{y}}^{\,-1}$}
\end{picture}
}
\put(80,0){\begin{picture}(24,24)
\put(0,0){\circle*{1}}
\qbezier[16](0,0)(16,0)(16,0)
\put(16,0){\circle*{1}}
\put(16,0){\line(0,1){16}}
\put(0,16){\circle*{1}}
\put(16,16){\circle*{1}}
\put(0,0){\vector(0,1){15.5}}
\put(-1,-3){\llap{$a$}}
\put(16.5,-3){$b$}
\put(-1,16.5){\llap{$d$}}
\put(16.5,14.5){$c$}
\put(-1,11){\llap{$\vec{\mathrm{z}}$}}
\put(0,16){\begin{picture}(24,8)(16,0)
\put(16,0){\vector(1,0){15.5}}
\put(32,0){\vector(1,1){8}}
\put(40,8){\vector(-1,0){16}}
\put(24,8){\vector(-1,-1){8}}
\put(24,-4){${\vec{\mathrm{x}}}$}
\put(37,2){$\vec{\mathrm{y}}$}
\put(32,9){${\vec{\mathrm{x}}}^{\,-1}$}
\put(21,2){$\vec{\mathrm{y}}^{\,-1}$}
\end{picture}
}
\end{picture}
}
\end{picture}
}
}\caption{The curvature mechanism.}\label{FigCurv}
\end{figure}
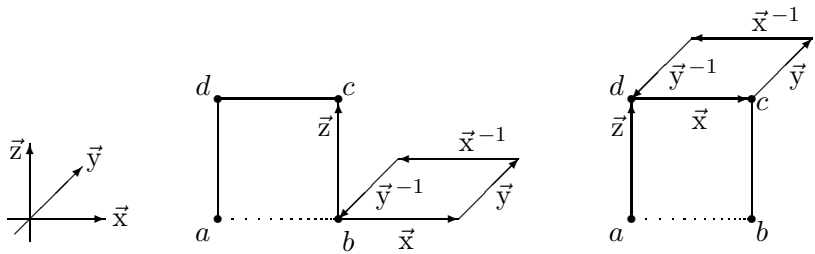

Let us inspect how the concept of Riemann curvature tensor works in the noncommutative setup when one transports an edge in 
the CW-\/complex's $1$-\/skeleton along a contour starting at a vertex formed by contracting an edge (see Fig.~\ref{FigCurv} on the cubic lattice).
Namely, let the edge $ab$ be contracted; consider the lattice element~${\vec{\mathrm{z}}}$. First, transport its starting point $a$ along the contour $\underrightarrow{{\vec{\mathrm{x}}}{\vec{\mathrm{y}}}{\vec{\mathrm{x}}}^{\,-1}{\vec{\mathrm{y}}}^{\,-1}}$ and \textsl{then} step along~${\vec{\mathrm{z}}}$; the walk's endpoint is~$c$. However, by walking the route~$\underrightarrow{{\vec{\mathrm{z}}} {\vec{\mathrm{x}}}{\vec{\mathrm{y}}}{\vec{\mathrm{x}}}^{\,-1}{\vec{\mathrm{y}}}^{\,-1}}$ and thus transporting the endpoint along the chosen contour, one reaches the vertex~$d$ instead of~$c$. By definition, a path connecting~$c$ to~$d$ is the value~$R_a({\vec{\mathrm{x}}},{\vec{\mathrm{y}}})\,{\vec{\mathrm{z}}}$ at~${\vec{\mathrm{z}}}$ of the quantum curvature operator for the path determined by the ordered pair~$({\vec{\mathrm{x}}},{\vec{\mathrm{y}}})$ at the point~$a$.

\begin{note}
In this text we postulate, not deriving the mass\/-\/energy balance equation $E=mc^2$ from the underlying geometric mechanism~\cite{Kiselev:Wald}, that 
a \textsl{presence} of a contracted edge is seen as mass, whereas a 
time\/-\/generating event of reconfiguration absorbs energy --creating mass-- by contracting an edge and releases that energy at the endpoints in the course of its decontraction;
this is the mass\/-\/energy correlation mechanism. (Note only that one may not measure the stored energy as ``force\/$\times$\/distance'' and thus introduce a stress of the lattice because length is undefined on it). 
\end{note}

\begin{corollary}
In absence of visible matter and energy, the vacuum can be curved and cause the force of gravity.
\end{corollary}

We expect that the \textsl{dark matter} is the configuration field of space contractions along its subsets; in fact, massive but invisible dark matter is not matter at all.

Let us define the \textsl{entropy}~$S$ of a given contraction of edges as minus the number of vertices~--- which themselves are not attached to the contracting edges but which neighbour via an edge with vertices that merge. Basically, this entropy of a contractions' configuration in space is the topologically\/-\/dual to \textsl{area} (here, the number of faces for the nearest surface that encapsulates the contracted edges).

For example, a contraction of one edge of the square lattice in~$\mathbb{E}^2$ produces a snowflake so that $S=-6$, making $2\times(-6)$ for two distant contractions. However, let us notice that the entropy of two consecutive edges for that tiling equals $-8$ and is equal to $-7$ for a corner. In view of this, a reconfiguration of contractions in a finite volume of the graph could be a thermodynamical process and gravity force could have the entropic origin, c.f.~\cite{Kiselev:EVerlinde2010}. 

We conjecture that 
initially, the entire space of the Universe was contracted to one point so that every tiling of it by the $1$-\/skeleton of the CW-\/complex was either that vertex or the vertex with tadpole attached to~it.
Simultaneously, we expect that space is topologically trivial so that all its possible tilings may not contain extra edges which would create shortcuts between distant cells.

\begin{remark}\label{Kiselev:RemLoll}
Numeric experiments~\cite{Kiselev:RLoll2009} reveal the following property which the CW\/-\/complex in~$\mathbb{E}^3$ gains in the course of bisection --i.e., making the tiling finer-- under an extra \textsl{ad hoc} assumption that the cells can be glued, orientation\/-\/preserving, along prescribed pairs of faces not necessarily to their true neighbours but possibly to sufficiently remote cells. This creates a possibility to obtain a topologically nontrivial CW\/-\/complex which nominally fills in~$\mathbb{E}^3$ but such that the local density of genus can be positive.
The probability of reconfigurations was postulated to drop exponentially with  increase of the $\mathbb{N}$-\/valued distance between cells.

Then the fundamental solution of the usual heat equation --or the square mean deviation of random walks-- was calculated by using a natural convention that the dissipating medium (e.g., smoke) or the random walks' endpoints spread freely through the faces of reconfigured tiling. The effective dimension was then determined from the rapidity of dissipation, and the modelling was repeated a suitable number of times.

Numeric experiment has shown that, as the sides of elementary domains become smaller but the effective distance, at which the probability of faces' reattachment drops $\exp(1)$ times, is kept constant, the effective dimension of $(3+1)$-\/dimensional combination of space and time drops from four to \textsl{exactly~two} in the continuous limit.\\
\centerline{\rule{1in}{0.7pt}}
\end{remark}

\noindent%
In the paper~\cite{Kiselev:Noel12} 
we try to view Physics as a text whose \textsl{meaning} is Nature.
We focus on its alphabet, glossary, 
grammar rules, and a possible location where the text is retrieved from, edited, and then stored back~to. 
We know that the text of Nature is incredibly interesting; in our efforts to read it, we have not yet advanced much in learning its grammar, and still more feebly we perceive the overall~plot.

\subsubsection*{Acknowledgements}
The author thanks the Organizing committee of 
the workshop `Group analysis of differential equations and integrable systems' (Protaras, Cyprus, 2012)
for partial support and a warm atmosphere during the meet\-ing.\quad %
%
%
This research was supported in part 
by 
JBI~RUG project~103511 (Groningen). 
A~part of this research was done while the author was visiting at 
the $\smash{\text{IH\'ES}}$ (Bures\/-\/sur\/-\/Yvette); 
the financial support and hospitality of this institution are gratefully acknowledged.

\LastPageEnding
\end{document}